\documentstyle[prd,preprint,tighten,aps,eqsecnum,%
amssymb,newlfont,epsfig]{revtex}
\begin{document}
\preprint{
\begin{tabular}{r}
UWThPh-2000-10\\
February 2000
\end{tabular}
}

\title{Gravitational anomalies in a dispersive approach\thanks{This work was partly
supported by Austria-Czech Republic Scientific collaboration, project KONTACT 1999-8.}}

\vspace{2cm}

\author{R.A. Bertlmann and E. Kohlprath\thanks{Supported by a Wissenschaftsstipendium der
Magistratsabteilung 18 der Stadt Wien.}}
\address{Institut f\"ur Theoretische Physik,
Universit\"at Wien\\
Boltzmanngasse 5,
A-1090 Vienna, Austria}

\maketitle

\vspace{2cm}

\begin{abstract}
The gravitational anomalies in two dimensions, specifically the Einstein anomaly and the
Weyl anomaly, are fully determined by means of dispersion relations. In this approach the
anomalies originate from the peculiar infrared feature of the imaginary part of the relevant
formfactor which approaches a $\delta$-function singularity at zero momentum squared when
$m \to 0$.

\end{abstract}

\section{Introduction}

Anomalies are the key to a deeper understanding of quantum field theory. Since their first
discovery by Adler \cite{Adler}, Bell and Jackiw \cite{BellJackiw}, and by Bardeen
\cite{Bardeen} they play a fundamental role in physics (for details see Ref.\cite{Bertlmann}).
Also in gravitation, where fermions interact with a gravitational field, anomalies may occur.
These are the Einstein anomaly -- signaling the breakdown of energy-momentum conservation --
the Lorentz anomaly -- reflecting an antisymmetric part of the energy-momentum tensor -- and
the Weyl anomaly -- expressing the nonvanishing tensor trace. We mainly refer to the work of
the authors \cite{CapperDuffNC} -- \cite{Leutwyler}
who have calculated (ultraviolet divergent) Feynman diagrams where the external
gravitational field couples to a fermion loop via the energy-momentum tensor.

In our paper we want to draw attention to an other approach with interesting features. It is
the dispersion relation (DR) approach which is an independent and complementary view of the
anomaly phenomenon as compared to the ultraviolet regularization procedures. In connection with
anomalies DR have been introduced by Dolgov and Zakharov
\cite{DolgovZakharov} and also by Kummer \cite{Kummer}. In the following several authors
\cite{HorejsiPRD} -- \cite{AdamBertlmannHofer1}
used successfully DR to determine the anomalies in the chiral current.
Recently Ho\v rej\v s\' \i{} and Schnabl \cite{HorejsiSchnabl} have applied the method to
the well-known trace anomaly which is
related to the broken dilatation (or scale) invariance. We extend in our work the method of
DR to the case of pure gravitation and perform the calculations in two dimensions.

\section{Structure of the amplitude}

The gravitational anomalies are determined by the one-loop diagram of a Weyl fermion
in a gravitational background field. Since it is sufficient to work with a linearized
gravitational field $g_{\mu\nu}=\eta_{\mu\nu}+\kappa h_{\mu\nu}$ and
$e^{a}_{\ \mu}=\eta^{a}_{\ \mu}+\frac{1}{2}\kappa h^{a}_{\ \mu}$,
where $e^{a}_{\,\,\mu}$ is the zweibein and $E_{a}^{\,\,\mu}$ its inverse
$E_{a}^{\,\,\mu}e^{a}_{\,\,\nu} = \delta^{\mu}_{\,\,\nu}$ \,  ,
we can start with the following linearized interaction Lagrangian 
(for convenience $\kappa$ is absorbed into $h^{a\mu}$, $\partial_{\mu}^{\psi}$ acts only
on $\psi$)

\begin{equation}
\mathcal{L}^{lin}_{I} = - \frac{i}{4} \left( h^{a\mu} \bar\psi \gamma_{a} \frac{1\pm\gamma_{5}}{2} \stackrel{\leftrightarrow}{\partial^{\psi}_{\mu}} \psi + h^{\mu}_{\ \mu} \bar\psi \gamma^{a} \frac{1\pm\gamma_{5}}{2} \stackrel{\leftrightarrow}{\partial^{\psi}_{a}} \psi \right)
 = -\frac{1}{2}  h_{\mu\nu} T^{\mu\nu} \, .
\end{equation}
From this expression follow the Feynman rules for the vertices in the loop
and the explicit form of the (symmetric) energy-momentum tensor

\begin{eqnarray} \label{energy-momentum-tensor}
T^{\mu\nu} &=& \frac{1}{2} \left( T^{\mu}_{\ a} E^{a\nu} + T^{\nu}_{\ a} E^{a\mu} \right) {}\nonumber\\{} &=& \frac{i}{4} \Biggl( \bar\psi E^{a\nu} \gamma_{a} \frac{1\pm\gamma_{5}}{2} \stackrel{\leftrightarrow}{\partial^{\mu}} \psi + \bar\psi E^{a\mu} \gamma_{a} \frac{1\pm\gamma_{5}}{2}  \stackrel{\leftrightarrow}{\partial^{\nu}} \psi \Biggr) \, ,
\end{eqnarray}
(note that in two dimensions the spin connection $\omega_{\mu}$ does not contribute,
see e.g. Ref.\cite{Bertlmann}).\\

Then the whole amplitude under consideration is given by the two-point function

\begin{equation} \label{amplitude expressed by T-product}
T_{\mu \nu \rho \sigma}(p) = i \int d^{2}x e^{i p x} \langle 0 \vert T \lbrack T_{\mu\nu}(x) T_{\rho\sigma}(0) \rbrack \vert 0 \rangle \, .
\end{equation}
Due to Lorentz covariance and symmetry we decompose and separate the amplitude in the
following way

\begin{equation}
T_{\mu \nu \rho \sigma}=T^{V}_{\mu \nu \rho \sigma}+T^{A}_{\mu \nu \rho \sigma}\label{formfactors1}
\end{equation}

\begin{eqnarray}
T^{V}_{\mu \nu \rho \sigma}(p) &=& p_{\mu}p_{\nu}p_{\rho}p_{\sigma} T_{1}(p^{2}) + (p_{\mu}p_{\nu} g_{\rho \sigma} + p_{\rho}p_{\sigma} g_{\mu \nu}) T_{2}(p^{2}) {} \nonumber \\ {} & &+ (p_{\mu}p_{\rho} g_{\nu \sigma} + p_{\mu}p_{\sigma} g_{\nu \rho} + p_{\nu}p_{\rho} g_{\mu \sigma} + p_{\nu}p_{\sigma} g_{\mu \rho}) T_{3}(p^{2}) {} \nonumber \\ {} & &+ g_{\mu \nu}g_{\rho \sigma} T_{4}(p^{2}) + (g_{\mu \rho}g_{\nu \sigma} + g_{\mu \sigma}g_{\nu \rho}) T_{5}(p^{2})\label{formfactors2}
\end{eqnarray}

\begin{eqnarray}
T^{A}_{\mu \nu \rho \sigma}(p) &=& (\varepsilon_{\mu \tau} p^{\tau} p_{\nu} p_{\rho} p_{\sigma} + \varepsilon_{\nu \tau} p^{\tau} p_{\mu} p_{\rho} p_{\sigma} + \varepsilon_{\rho \tau} p^{\tau} p_{\mu} p_{\nu} p_{\sigma} + \varepsilon_{\sigma \tau} p^{\tau} p_{\mu} p_{\nu} p_{\rho}) T_{6}(p^{2}) {} \nonumber \\ {}& & + (\varepsilon_{\mu \tau} p^{\tau} p_{\nu} g_{\rho \sigma} + \varepsilon_{\nu \tau} p^{\tau} p_{\mu} g_{\rho \sigma} + \varepsilon_{\rho \tau} p^{\tau} p_{\sigma} g_{\mu \nu} + \varepsilon_{\sigma \tau} p^{\tau} p_{\rho} g_{\mu \nu}) T_{7}(p^{2}) {} \nonumber \\ {}& &+ \Bigl[\varepsilon_{\mu \tau} p^{\tau} (p_{\rho} g_{\nu \sigma} + p_{\sigma} g_{\nu \rho}) + \varepsilon_{\nu \tau} p^{\tau} (p_{\rho} g_{\mu \sigma} + p_{\sigma} g_{\mu \rho})  {} \nonumber \\ {}& & +\varepsilon_{\rho \tau} p^{\tau} (p_{\mu} g_{\nu \sigma} + p_{\nu} g_{\mu \sigma}) + \varepsilon_{\sigma \tau} p^{\tau} (p_{\mu} g_{\nu \rho} + p_{\nu} g_{\mu \rho})\Bigr] T_{8}(p^{2}) \, .\label{formfactors3}
\end{eqnarray}
The functions $T_{1}(p^{2}), ... ,T_{8}(p^{2})$ are the formfactors that are to be
evaluated.\\

The classical properties of the energy-momentum tensor ($T_{\mu\nu} = T_{\nu\mu}$ symmetric,
$\nabla^{\mu}T_{\mu\nu} =~0$ conserved, $T^{\mu}_{\ \mu} = 0$ traceless)
lead to the following canonical (naive) Ward identities:
\begin{enumerate}
\item $T_{\mu\nu\rho\sigma}(p) = T_{\nu\mu\rho\sigma}(p)$
\item $p^{\mu}T_{\mu\nu\rho\sigma}(p) = 0$
\item $g^{\mu\nu}T_{\mu\nu\rho\sigma}(p) = 0$ \,.
\end{enumerate}

We are interested in the pure Einstein anomaly therefore we demand the quantized
energy-momentum tensor to be symmetric, which is always possible to achieve. Thus the symmetry
property 1.) of the amplitude is fulfilled, however, the Ward identity (WI) 2.) and the trace
identity (TI) 3.) need not be satisfied, they can be broken by the Einstein- and the Weyl
anomaly respectively.

The canonical Ward identities we re-express by the formfactors. For the pure tensor part of
the amplitude the WI may be written as

\begin{eqnarray}
&&p^{2} T_{1} + T_{2} + 2 T_{3} = 0 \label{VWI1} \\
&&p^{2} T_{2} + T_{4} = 0 \label{VWI2} \\
&&p^{2} T_{3} + T_{5} = 0 \, , \label{VWI3}
\end{eqnarray}
and the TI as

\begin{eqnarray}
&&p^{2} T_{1} + 2 T_{2} + 4 T_{3} = 0 \, . \label{TI1}
\end{eqnarray}

In the following we shall use a renormalization procedure which keeps the WI in the pure
tensor part (\ref{VWI1}) -- (\ref{VWI3}) so that the anomaly occurs only in the pseudotensor
part of the amplitude.

For convenience we split the pure tensor piece of the loop in the following way

\begin{equation} \label{pure vector amplitude}
T^{V}_{\mu \nu \rho \sigma}(p) = \frac{1}{2} T^{pv}_{\mu \nu \rho \sigma}(p) - T^{dv}_{\mu \nu \rho \sigma}(p) \, ,
\end{equation}
where $T^{pv}_{\mu\nu\rho\sigma}$ represents the loop with the identity instead of the
chirality projectors and $T^{dv}_{\mu\nu\rho\sigma}$ denotes the part proportional
to $m^{2}$.

Finally, the axial part of the amplitude is connected to the vector part due to
relation

\begin{equation} \label{elimination of gamma5}
\gamma_{\mu}\gamma_{5} = -\varepsilon_{\mu \nu} \gamma^{\nu}
\end{equation}
(valid only in 2 dimensions) where our conventions are $g_{00} = - g_{11} = 1$,
$\varepsilon^{01} = 1$ and $\gamma^{0} = \sigma^{2}$, $\gamma^{1} = i\sigma^{1}$,
$\gamma^{5} = \gamma^{0}\gamma^{1} =  \sigma^{3}$, and $\sigma^{i}$ being the Pauli matrices.

\section{Dispersion relations}

The formfactors of the amplitude (\ref{formfactors1}) -- (\ref{formfactors3}) can be expressed
by dispersion relations which relate the real part of the amplitude to its imaginary part.
The imaginary parts of the amplitude can be easily calculated via Cutkosky's rule
\cite{Cutkosky}. In this way we find the imaginary parts of all formfactors in the total
amplitude $T_{\mu\nu\rho\sigma}$ (for details see Ref. \cite{Kohlprath}):

\begin{eqnarray}
Im T_{1}(p^{2}) &=& -\frac{1}{4} J_{0}\frac{m^{2}}{p^{2}}\left(1-4\frac{m^{2}}{p^{2}}\right) \\
Im T_{2}(p^{2}) &=& -\frac{1}{48} J_{0}\,p^{2}\left(1-8\frac{m^{2}}{p^{2}}+16\frac{m^{4}}{p^{4}}\right) \\
Im T_{3}(p^{2}) &=& \frac{1}{96} J_{0}\,p^{2}\left(1+\frac{m^{2}}{p^{2}}-20\frac{m^{4}}{p^{4}}\right) \\
Im T_{4}(p^{2}) &=& \frac{1}{48} J_{0}\,p^{4}\left(1-8\frac{m^{2}}{p^{2}}+16\frac{m^{4}}{p^{4}}\right) \\
Im T_{5}(p^{2}) &=& -\frac{1}{96} J_{0}\,p^{4}\left(1-2\frac{m^{2}}{p^{2}}-8\frac{m^{4}}{p^{4}}\right) \\
Im T_{6}(p^{2}) &=& \pm \frac{1}{16} J_{0}\frac{m^{2}}{p^{2}}\left(1-4\frac{m^{2}}{p^{2}}\right) \\
Im T_{7}(p^{2}) &=& \pm \frac{1}{192} J_{0}\,p^{2}\left(1-8\frac{m^{2}}{p^{2}}+16\frac{m^{4}}{p^{4}}\right) \\
Im T_{8}(p^{2}) &=& \mp \frac{1}{384} J_{0}\,p^{2}\left(1+4\frac{m^{2}}{p^{2}}-32\frac{m^{4}}{p^{4}}\right)
\end{eqnarray}
with the threshold function

\begin{equation}
J_{0} = \frac{1}{p^{2}}\left(1-\frac{4m^{2}}{p^{2}}\right)^{-1/2}\theta(p^{2}-4m^{2}) \, .
\end{equation}
Considering on the other hand the amplitude $T^{pv}_{\mu\nu\rho\sigma}$ we get the following
imaginary parts:

\begin{eqnarray}
Im A_{1}(p^{2}) &=& -\frac{1}{2} J_{0}\frac{m^{2}}{p^{2}}\left(1-4\frac{m^{2}}{p^{2}}\right) \label{Im A1}\\
Im A_{2}(p^{2}) &=& -\frac{1}{24} J_{0}\,p^{2}\left(1-8\frac{m^{2}}{p^{2}}+16\frac{m^{4}}{p^{4}}\right) \\
Im A_{3}(p^{2}) &=& \frac{1}{48} J_{0}\,p^{2}\left(1+4\frac{m^{2}}{p^{2}}-32\frac{m^{4}}{p^{4}}\right) \\
Im A_{4}(p^{2}) &=& \frac{1}{24} J_{0}\,p^{4}\left(1-8\frac{m^{2}}{p^{2}}+16\frac{m^{4}}{p^{4}}\right) \\
Im A_{5}(p^{2}) &=& -\frac{1}{48} J_{0}\,p^{4}\left(1+4\frac{m^{2}}{p^{2}}-32\frac{m^{4}}{p^{4}}\right). \label{Im A5}
\end{eqnarray}
Clearly, the imaginary parts (\ref{Im A1}) -- (\ref{Im A5}) of the amplitude
$T^{pv}_{\mu\nu\rho\sigma}$ satiesfy the WI (\ref{VWI1})--(\ref{VWI3}) with
$T_{i} \to Im A_{i}(p^{2})$, and the subtraction procedure we choose in the following
keeps this property for the entire formfactors $A_{i}(p^{2})$.\\

Now we start with an unsubtracted dispersion relation for the formfactors

\begin{equation}\label{unsubtracted DR}
T(p^{2}) = \frac{1}{\pi} \int\limits^{\infty}_{4m^{2}}\!\!\frac{dt}{t-p^{2}} Im T(t)
\end{equation}
and we observe that, for instance, the integral for $T_{1}(p^{2})$ is convergent
whereas for $T_{2}(p^{2})$ it is logarithmically divergent and needs to be subtracted
once, and for $T_{4}(p^{2})$ it is linearly divergent and needs to be subtracted
twice. We can infer already from the $p^{2} = t$ behaviour of the imaginary parts which
kind of dispersion relation we have to use.

So for the formfactors $T_{1}, T_{6}, A_{1}$ an unsubtracted DR is sufficient and
we get

\begin{eqnarray}
T_{1}(p^{2}) &=& \mp 4 T_{6}(p^{2}) = \frac{1}{2}A_{1}(p^{2}) {}\nonumber\\{}&=& -\frac{1}{4\pi} \int\limits^{\infty}_{4m^{2}}\!\!\frac{dt}{t-p^{2}} \frac{m^2}{t^2}\left(1-\frac{4m^{2}}{t}\right)^{ \hspace{-3pt}\frac{1}{2}} {}\nonumber\\{}&=& \frac{1}{p^{2}}\left[\frac{1}{24\pi}  - \frac{1}{2\pi} \frac{m^{2}}{p^{2}} + \frac{1}{2\pi} \frac{m^{2}}{p^{2}} a(p^{2})\right]
\end{eqnarray}
with

\begin{equation} \label{definition of a}
a(p^{2}) = \sqrt{\frac{4m^{2}-p^{2}}{p^{2}}} \arctan \sqrt{\frac{p^{2}}{4m^{2}-p^{2}}} \enspace .
\end{equation}
A once subtracted DR defined by

\begin{equation}
T^{R}(p^{2}) = T(p^{2}) - T(0) = \frac{p^{2}}{\pi} \int\limits^{\infty}_{4m^{2}}\!\!\frac{dt}{t-p^{2}} \frac{1}{t}Im T(t)
\end{equation}
we use for the following formfactors

\begin{eqnarray}
T^{R}_{2}(p^{2}) &=& \mp 4 T^{R}_{7}(p^{2}) = \frac{1}{2}A^{R}_{2}(p^{2}) {}\nonumber\\{}&=& \frac{p^{2}}{\pi} \int\limits^{\infty}_{4m^{2}}\!\!\frac{dt}{t-p^{2}} \frac{1}{t^{2}}\left(1-\frac{4m^{2}}{t}\right)^{\hspace{-3pt}-\frac{1}{2}} \hspace{-3pt} \left(- \frac{1}{48}t + \frac{1}{6}m^{2}-\frac{1}{3}\frac{m^{4}}{t}\right) {}\nonumber\\{}&=& -\frac{1}{18\pi} + \frac{1}{6\pi} \frac{m^{2}}{p^{2}} + \frac{1}{24\pi} \Bigl(1- 4 \frac{m^{2}}{p^{2}} \Bigr) a(p^{2})\\
T^{R}_{3}(p^{2}) &=& \frac{p^{2}}{\pi} \int\limits^{\infty}_{4m^{2}}\!\!\frac{dt}{t-p^{2}} \frac{1}{t^{2}}\left(1-\frac{4m^{2}}{t}\right)^{\hspace{-3pt}-\frac{1}{2}} \hspace{-3pt} \left( \frac{1}{96}t + \frac{1}{96}m^{2}-\frac{5}{24}\frac{m^{4}}{t}\right) {}\nonumber\\{}&=& \frac{7}{576\pi} + \frac{5}{48\pi} \frac{m^{2}}{p^{2}} - \frac{1}{48\pi} \Bigl( 1 + 5 \frac{m^{2}}{p^{2}} \Bigr) a(p^{2})\\
A^{R}_{3}(p^{2}) &=& \frac{p^{2}}{\pi} \int\limits^{\infty}_{4m^{2}}\!\!\frac{dt}{t-p^{2}} \frac{1}{t^{2}}\left(1-\frac{4m^{2}}{t}\right)^{\hspace{-3pt}-\frac{1}{2}} \hspace{-3pt} \left( \frac{1}{48}t + \frac{1}{12}m^{2}-\frac{2}{3}\frac{m^{4}}{t}\right) {}\nonumber\\{}&=& \frac{1}{72\pi} + \frac{1}{3\pi} \frac{m^{2}}{p^{2}} - \frac{1}{24\pi} \Bigl( 1 + 8 \frac{m^{2}}{p^{2}} \Bigr) a(p^{2})\\
T^{R}_{8}(p^{2}) &=& \mp \frac{p^{2}}{\pi} \int\limits^{\infty}_{4m^{2}}\!\!\frac{dt}{t-p^{2}} \frac{1}{t^{2}}\left(1-\frac{4m^{2}}{t}\right)^{\hspace{-3pt}-\frac{1}{2}} \hspace{-3pt} \left( \frac{1}{384}t + \frac{1}{96}m^{2}-\frac{1}{12}\frac{m^{4}}{t}\right) {}\nonumber\\{}&=&  \mp \frac{1}{576\pi} \mp \frac{1}{24\pi} \frac{m^{2}}{p^{2}} \pm \frac{1}{192\pi} \Bigl( 1 + 8 \frac{m^{2}}{p^{2}} \Bigr) a(p^{2}) \, .
\end{eqnarray}
For the remaining formfactors a twice subtracted DR defined by

\begin{equation}
T^{R}(p^{2}) = T(p^{2}) - T(0) -p^{2} \left. \frac{d}{dp^{2}}T(p^{2})\right|_{p^{2}=0} = \frac{p^{4}}{\pi} \int\limits^{\infty}_{4m^{2}}\!\!\frac{dt}{t-p^{2}} \frac{1}{t^{2}}Im T(t)
\end{equation}
is necessary and we find

\begin{eqnarray}
T^{R}_{4}(p^{2}) &=& \frac{1}{2}A^{R}_{4}(p^{2}) = \frac{p^{4}}{\pi} \int\limits^{\infty}_{4m^{2}}\!\!\frac{dt}{t-p^{2}} \frac{1}{t^{3}}\left(1-\frac{4m^{2}}{t}\right)^{\hspace{-3pt}-\frac{1}{2}} \hspace{-3pt} \left( \frac{1}{48}t^{2} - \frac{1}{6}t m^{2}+\frac{1}{3} m^{4} \right) {}\nonumber\\{}&=& p^{2}\left[\frac{1}{18\pi} - \frac{1}{6\pi}\frac{m^{2}}{p^{2}} - \frac{1}{24\pi} \Bigl(1-4\frac{m^{2}}{p^{2}} \Bigr) a(p^{2})\right]\\
T^{R}_{5}(p^{2}) &=& \frac{p^{4}}{\pi} \int\limits^{\infty}_{4m^{2}}\!\!\frac{dt}{t-p^{2}} \frac{1}{t^{3}}\left(1-\frac{4m^{2}}{t}\right)^{\hspace{-3pt}-\frac{1}{2}} \hspace{-3pt} \left( -\frac{1}{96}t^{2} + \frac{1}{48}t m^{2}+\frac{1}{12} m^{4} \right) {}\nonumber\\{}&=& p^{2}\left[-\frac{5}{288\pi} - \frac{1}{24\pi} \frac{m^{2}}{p^{2}} + \frac{1}{48\pi} \Bigl(1+ 2 \frac{m^{2}}{p^{2}} \Bigr) a(p^{2})\right]\\
A^{R}_{5}(p^{2}) &=& \frac{p^{4}}{\pi} \int\limits^{\infty}_{4m^{2}}\!\!\frac{dt}{t-p^{2}} \frac{1}{t^{3}}\left(1-\frac{4m^{2}}{t}\right)^{\hspace{-3pt}-\frac{1}{2}} \hspace{-3pt} \left( -\frac{1}{48}t^{2} - \frac{1}{12}t m^{2}+\frac{2}{3}m^{4}\right) {}\nonumber\\{}&=& p^{2}\left[-\frac{1}{72\pi} - \frac{1}{3\pi} \frac{m^{2}}{p^{2}} + \frac{1}{24\pi} \Bigl(1+ 8 \frac{m^{2}}{p^{2}} \Bigr) a(p^{2})\right] \, .
\end{eqnarray}

With these explicit expressions for the formfactors we have determined the whole amplitude
$T_{\mu\nu\rho\sigma}$, Eqs.(\ref{amplitude expressed by T-product})--(\ref{formfactors3}),
from which the correct Ward identities will follow.

\section{Anomalous Ward identities and gravitational anomalies}

Now we turn to the calculation of the Ward identities and gravitational anomalies. We consider
the massless limit, $m \to 0$, where the formfactors $2T_{i} \to A_{i}$ (i = 1,...,5) fulfill
the WI (\ref{VWI1}) -- (\ref{VWI3}). This means that the WI for the pure tensor part is
satiesfied
\begin{equation}
p^{\mu}T^{V}_{\mu\nu\rho\sigma}(p)=0 \, .
\end{equation}
Next we calculate the WI for the pseudo tensor part. We use the formfactor identities
\begin{equation} \label{axial--vector}
T_{6}=\mp \frac{1}{4}T_{1}\, ,\qquad T_{7}=\mp\frac{1}{4}T_{2}\, , \qquad T_{8}=\mp\frac{1}{4}T_{3}\, ,
\end{equation}
and we obtain the anomalous result
\begin{equation} 
p^{\mu} T^{A}_{\mu \nu \rho \sigma}(p) = \mp \frac{1}{4} p^{2}T_{1}\ \varepsilon_{\nu\tau}p^{\tau}(p_{\rho}p_{\sigma}-g_{\rho\sigma}p^{2}) \, . \label{pTAa}
\end{equation}
The anomalous WI depends only on the finite formfactor $T_{1} = \mp 4 T_{6}$ with its explicit
result
\begin{eqnarray}
T_{1}(p^{2}) &=& \mp 4\ T_{6}(p^{2}) = \frac{1}{24\pi p^{2}} \, . \label{T1 for WI}
\end{eqnarray}
So the anomaly is independent of a specific renormalization procedure (as long as it preserves
the WI (\ref{VWI1})-(\ref{VWI3})) and we agree with the anomaly results of Tomiya \cite{Tomiya}
and Alvarez-Gaum\'e and Witten \cite{AlvarezWitten} who applied quite different
(regularization) methods.

We want to emphasize that our subtraction procedure is {\em the\/} `natural' choice dictated
by the $t-$behaviour of the imaginary parts $ImT_{i}(t)$ of the formfactors, which automatically
shifts the total anomaly into the pseudotensor part of the WI (\ref{pTAa}).\\

What is the origin of the anomaly in this dispersive approach? The source of the anomaly
is the existence of a superconvergence sum rule for the imaginary part of the formfactor
$T_{1}(p^{2})$
\begin{equation}
\int\limits^{\infty}_{0}\!dt\,Im\,T_{1}(t)=-\frac{m^2}{4}\int\limits^{\infty}_{4m^{2}}\!\!\frac{dt}{t^{2}}\left(1-\frac{4m^{2}}{t}\right)^{\hspace{-3pt}\frac{1}{2}}=-\frac{1}{24} \enspace .
\end{equation} 
The anomaly originates from a $\delta$-function singularity of $ImT_{1}(t)$ when the
threshold $t = 4m^{2} \to 0$ approaches zero (the infrared region)
\begin{equation}
\lim_{m\rightarrow 0} Im\,T_{1}(t)=-\lim_{m\rightarrow 0}\frac{m^{2}}{4t^{2}}\left(1-\frac{4m^{2}}{t}\right)^{\hspace{-3pt}\frac{1}{2}}\theta(t-4m^{2})=-\frac{1}{24}\delta(t) \, .
\end{equation}
The limit must be performed in a distributional sense.\\
Then the unsubtracted dispersion relation for $T_{1}(p^{2})$, Eq.(\ref{unsubtracted DR}), 
provides the result (\ref{T1 for WI}).
This threshold singularity of the imaginary part of the relevant formfactor is a typical
feature of the DR approach for calculating the anomaly
(see e.g. Refs.\cite{Bertlmann}, \cite{HorejsiPRD} --  \cite{HorejsiSchnabl}).\\

Next we turn to the energy-momentum tensor. From the anomalous WI (\ref{pTAa})
we can deduce the linearized consistent Einstein (or diffeomorphism) anomaly
\begin{equation} \label{linearized consistent Einstein anomaly}
\partial^{\mu}\langle T_{\mu\nu} \rangle = \mp \frac{1}{192\pi} \varepsilon_{\mu\nu} \partial^{\mu} \left( \partial_{\alpha}\partial_{\beta} h^{\alpha\beta} - \partial_{\alpha}\partial^{\alpha}h^{\beta}_{\ \beta}\right).
\end{equation}
Result (\ref{linearized consistent Einstein anomaly}) is indeed the linearization of the
exact result that follows from differential geometry and topology (see for instance
Ref.\cite{Bertlmann}).

Now what about the covariant Einstein anomaly? It arises when considering the covariantly
transforming energy-momentum tensor $\tilde T_{\mu\nu}$ which is related to our tensor
definition (\ref{energy-momentum-tensor}) by the Bardeen-Zumino polynomial $\mathcal{P}_{\mu\nu}$ \cite{BardeenZumino}
\begin{equation}
\langle \tilde T_{\mu\nu}\rangle = \langle T_{\mu\nu}\rangle + \mathcal{P}_{\mu\nu} \, .
\end{equation}
This polynomial is calculable explicitly, for its linearized version we find
\begin{equation}
\partial^{\mu}\mathcal{P}_{\mu\nu} = \mp \frac{1}{192\pi} \varepsilon_{\mu\nu} \partial^{\mu} \left( \partial_{\alpha}\partial_{\beta} h^{\alpha\beta} - \partial_{\alpha}\partial^{\alpha}h^{\beta}_{\ \beta}\right)
\end{equation}
leading to the linearized covariant Einstein anomaly
\begin{equation}
\partial^{\mu}\langle\tilde T_{\mu\nu} \rangle = \mp \frac{1}{96\pi} \varepsilon_{\mu\nu} \partial^{\mu} \left( \partial_{\alpha}\partial_{\beta} h^{\alpha\beta} - \partial_{\alpha}\partial^{\alpha}h^{\beta}_{\ \beta}\right).
\end{equation}
It is twice the linearized consistent result (\ref{linearized consistent Einstein anomaly})
as it should be.\\

Finally we also calculate the trace identity. Using again relations (\ref{axial--vector})
and taking into account the WI (\ref{VWI1}) provides us the anomalous result
\begin{equation}
T^{\mu}_{\ \mu\rho\sigma}=-p^{2}T_{1}\Biggl[\left(p_{\rho}p_{\sigma}-p^{2}g_{\rho\sigma}\right) \mp\frac{1}{4}\left( \varepsilon_{\rho\lambda}p^{\lambda}p_{\sigma}+\varepsilon_{\sigma\lambda}p^{\lambda}p_{\rho}\right)\Biggl] \, .
\end{equation}
Also the anomalous TI depends only on the finite formfactor $T_{1} = \mp 4 T_{6}$
so that it is independent of a specific renormalization procedure (which preserves
the WI (\ref{VWI1})-(\ref{VWI3})).

Inserting the formfactor, Eq.(\ref{T1 for WI}), implies the following linearization of the
Weyl (or trace) anomaly
\begin{equation} \label{linearized Weyl anomaly}
\langle T^{\mu}_{\ \mu} \rangle = \frac{1}{48\pi}\Biggl[ \left(\partial_{\mu}\partial_{\nu}h^{\mu\nu}-\partial_{\mu}\partial^{\mu}h^{\nu}_{\ \nu}\right) \mp\frac{1}{2}\varepsilon_{\mu\lambda}\partial^{\lambda}\partial_{\nu}h^{\mu\nu}\Biggr] \, .
\end{equation}
Again, result (\ref{linearized Weyl anomaly}) is indeed the linearization of the exact result
(see for instance Ref.\cite{Bertlmann}).

Adding last but not least the Bardeen-Zumino polynomial $\mathcal{P}_{\mu\nu}$
with its linearization
\begin{equation}
\mathcal{P}^{\mu}_{\ \mu}=\pm \frac{1}{96\pi}\varepsilon^{ab}\partial^{\mu}\partial_{b}h_{\mu a}
\end{equation}
we find for the linearized covariant trace anomaly
\begin{equation}
\langle T^{\mu}_{\ \mu} \rangle = \frac{1}{48\pi}\left( \partial_{\mu}\partial_{\nu}h^{\mu\nu}-\partial_{\mu}\partial^{\mu}h^{\nu}_{\ \nu}\right) \, .
\end{equation}
Clearly this result is in agreement with Ref.\cite{Leutwyler}.

\section{Conclusions}

We have investigated an alternative method, the DR approach, to calculate the gravitational
anomalies. The method appears quite appealing, all one has to calculate is the imaginary part
of just one formfactor of the amplitude $T_{\mu\nu\rho\sigma}(p)$, namely $Im T_{1}(p^{2})$,
which is an easy task.

Our `natural' subtraction procedure for the formfactors implies that the pure tensor WI
(\ref{VWI1}) -- (\ref{VWI3}) for the renormalized formfactors is satisfied (in the limit
$m \to 0$), so that the total anomaly is automatically shifted into the
pseudotensor part of the WI (\ref{pTAa}). From the anomalous WI and the anomalous TI
follow the linearized Einstein- and Weyl anomaly, and we have also determined their covariant
versions. An analogous dispersive calculation of the anomalous commutators of the
energy-momentum tensors -- the gravitational Schwinger terms -- is given elsewhere
\cite{BertlmannKohlprathST}. 

The anomalies originate from the peculiar infrared feature of the imaginary part of the
relevant formfactor $T_{1}(p^{2})$ which approaches a $\delta$-function singularity at zero
momentum squared when $m \to 0$.

We have considered the anomalies in two dimensions, where the essential features of the
DR approach already show up and all calculations come out very simple. However, this
convenient computational simplicity is a very special feature of the two space-time
dimensions, in higher dimensions the calculations will turn out much more complicated.

\end{document}